 \documentclass[%
reprint,
amsmath,amssymb,
aps, 
]{revtex4-2}

\usepackage{amsmath,amssymb}
\usepackage[dvipdfmx]{graphicx}% Include figure files
\usepackage{hyperref}% add hypertext capabilities
\usepackage{physics}
\newcommand{\ee}{{\mathrm{e}}}
\newcommand{\ri}{{\mathrm{i}}}

\begin{document}

\title{Ensemble model of turbulence based on states of constant flux in wavenumber space}

\author{Kyo Yoshida}
\affiliation{%
Division of Physics, Faculty of Pure and Applied Sciences, University of Tsukuba
}
\date{\today}

%%%%%%\maketitle

\begin{abstract}
An ensemble model of turbulence is proposed.  
The ensemble consists of flow fields 
in which the flux of an inviscid conserved quantity, 
such as energy (or enstrophy in two-dimensional flow fields), 
across the wavenumber $k$ is a constant independent of $k$ 
in an appropriate range. 
Two-dimensional flow fields of constant enstrophy flux are 
sampled randomly by a Monte Carlo method.  
The energy spectra $E_k$ of the flow fields are consistent 
with the scaling $E_k \propto k^{-3}\qty(\ln(k/k_\mathrm{b}))^{-1/3}$ 
where $k_\mathrm{b}$ is the bottom wavenumber of the constant-flux range.  
\end{abstract}

\maketitle

\section{Introduction}
%\paragraph{Introduction.}
The motions of viscous fluids can be modeled by the Navier-Stokes (NS) 
equation.  
When the Reynolds number $Re:=UL/\nu$, where $U$ and $L$ are typical velocity 
and length, respectively, of the flow and $\nu$ is the kinematic viscosity, 
is very large, the individual solutions are sensitive to small disturbances 
and appear to be irregular in space and time.  
Motions of fluid in such a situation are called turbulence.  
It seems natural to employ the concept of statistics or probability 
in considering turbulence.  However, the statistical theory of turbulence 
is far from being established as we discuss below.  

Let us first recall the equilibrium statistical mechanics for the comparison.  
The establishment of the statistical mechanics owes essentially to 
the ensemble picture which was first introduced in the consistent 
formalism by Gibbs\cite{Gibbs1902}.  See, e.g., Ref.\cite{Inaba2021} 
for a historical review. 
In the ensemble picture, a macroscopic state is modeled by 
an ensemble of microscopic states.  Macroscopic quantities can be derived 
from the averages of corresponding microscopic quantities over the ensemble.  
For thermal equilibrium states, the ensemble models can be microcanonical, 
canonical, or grand canonical, and they can be defined by the Hamiltonian 
of the system and corresponding thermodynamic variables.  
Recently, the ensemble picture has been reviewed based on typicality 
and in the context of the thermalization of isolated quantum systems.  
See, e.g., Ref.\cite{Tasaki2016} and the references therein.  
Although an alternative formalism based on so-called individualist views 
has been discussed especially for taking nonequilibrium processes into 
consideration\cite{GoldsteinLebowitzTumulkaZanghi2020,Yoshida2020}, 
the ensemble picture provides the complete, concise, and most 
feasible computational tool of the statistical mechanics 
at least as far as thermal equilibrium states are concerned. 

Since turbulence is a nonequilibrium state in the sense that 
there is a macroscopic flow of energy coming into the system 
by external forces and going out by viscosity, 
the ensemble model for thermal equilibrium states such as 
the microcanonical or canonical ensemble model can not be applied.  
A mathematically rigorous choice of the ensemble is 
a stationary probability measure on the state space.  
The analysis related to the stationary probability measure 
for the NS turbulence is quite difficult, 
but see Ref.~\cite{Bedrossian2022batchelor} for 
a recent related analysis on the passive scalar turbulence. 
There is an idea of representing the stationary probability measure 
by periodic orbits.  See, e.g., Ref. \cite{KawaharaKida2001}.  
The periodic orbits were searched numerically, however, 
the search becomes hard with the increase of the Reynolds number.  
When we consider the external forcing as a random field in space and time, 
the probability measure is attributed to trajectories in the state space. 
The Martin-Siggia-Rose-Janssen-de Dominicis procedure
\cite{MartinSiggiaRose1973,Janssen1976,deDominicis1976} may be used to 
treat the problem in a field-theoretic formalism.  Recently, 
a nonperturbative renormalization group analysis has been attempted 
within the formalism\cite{CanetDelamotteWschebor2016}.  
The entropy method (EM) is one of the methods that treat 
the probability measure on the state space in 
an explicit manner\cite{EdwardsMcComb1969}. 
The relation between EM and the model in this study will be discussed in 
Sec.~\ref{sec:discussions}.  
In statistical closure approaches, one abandons the idea of specifying 
the ensemble of states or trajectories and resorts to derive closed 
relations between low order moments upon some assumptions for 
the approximation.  Especially, Lagrangian spectral (two-point) closures 
such as the abridged Lagrangian history direct interaction 
approximation\cite{Kraichnan1965} and the Lagrangian renormalized 
approximation (LRA)\cite{Kaneda1981} are capable of deriving 
the Kolmogorov spectrum up to the estimate of the universal constant.  
%More rough closure approximations with fitting parameters may be employed 
%for engineering and geophysical applications.  
See Ref.~\cite{Zhou2021} for a comprehensive review of the statistical closure 
approaches.  Although turbulence has been studied from various aspects, 
it may be said that there is no established statistical theory of turbulence 
that can compare with the ensemble models in the equilibrium statistical 
mechanics as of now.  

In this study, we propose a new ensemble model of turbulence 
expecting its potential to be one of the effective tools for 
the statistical theory of turbulence.  
The model incorporates the concept of cascade 
at the level of its construction.  
%Here, the conserved quantities mean those that are conserved 
%under the nonlinear interactions, such as the energy, and 
Here, the cascade means successive local transfers of 
an inviscid conserved quantity from large scales to small scales 
or vice versa.  In the case of three-dimensional turbulence, 
the energy cascades from large scales to small scales as 
described by Richardson~\cite{inRichardson1922} as early as 1922. 
When the turbulence is at a stationary state in a statistical sense, 
the mean energy injection and dissipation rate, 
$\epsilon_\mathrm{in}$ and $\epsilon_\mathrm{d}$ respectively, equilibrate 
and the energy flows with a constant flux independent of the scale $\ell$, 
i.e., $\Phi_\ell=\epsilon_\mathrm{in}=\epsilon_\mathrm{d}$, 
in the intermediate scale range so-called the inertial range, where 
$\Phi_\ell$ is the energy flux from the scales larger than $\ell$ 
to those smaller than $\ell$.  The notion of universality in the 
turbulence statistics is that statistical quantities in the inertial 
range are irrelevant to the details of forcing and dissipation outside 
the range when the inertial range is sufficiently broad, i.e., the Reynolds 
number is very large.  Kolmogorov's hypotheses of similarity claim 
that the mean energy dissipation rate $\epsilon_\mathrm{d}$ is the only relevant 
parameter\cite{Kolmogorov1941a}.  
Although the hypotheses have been denied in the context of intermittency 
(see, e.g., Ref.~\cite{Frisch1995}), the significance of the parameter 
$\epsilon_\mathrm{d}$ still remains.  
Since $\epsilon_\mathrm{d}$ is a quantity associated with 
the small scales where the viscosity is dominant, it may be appropriate 
to put the energy flux $\Phi_\ell$, which is a quantity associated with 
the scales in the inertial range, at the center of the construction 
of the model and consider $\epsilon_\mathrm{d}$ as an external parameter.  
Note that it was pointed out by Onsager~\cite{Onsager1949} that the energy 
dissipation could take place in the absence of viscosity and 
the modern analysis of the issue essentially involves the energy flux 
$\Phi_\ell$.  See, e.g., Ref.\cite{Eyink2018}.  
In this study, we formulate the ensemble model of states whose energy flux 
is constant, i.e., $\Phi_\ell=\epsilon_\mathrm{d}$, for the scales $\ell$ in 
the inertial range.  The formulation is given in the wavevector space.  

\section{Setting of the system}
%\noindent
%\paragraph{Setting of the system.}
We consider an incompressible fluid in a $d$-dimensional domain 
$[0,L]^d$ with periodic boundary conditions, where $d\ge 2$ and 
usually $d=3$.  A state, symbolically denoted by $\vb*{u}$, 
of the fluid is specified by an incompressible velocity vector field. 
Let $\vb*{u}_{\vb*{k}}:=(2\pi)^{-d} \int_{[0,L]^d}\dd{\vb*{x}} 
\ee^{-\ri \vb*{k}\cdot\vb*{x}}\vb*{u}(\vb*{x}) 
(\vb*{k}\in\mathcal{K})$ denotes the Fourier 
coefficients of the velocity field 
where $\mathcal{K}$ is a set of wavevectors 
$\mathcal{K}:=\{(k_1,\ldots,k_d)|k_j=m\Delta k, m\in \mathbb{Z}, 
k< k_{\max}\}-\{\vb*{0}\}$, $k:=|\vb*{k}|$, $\Delta k:=2\pi/L$ 
and the cutoff wavenumber $k_{\max}$ is introduced. 
The reality of $\vb*{u}$ in the physical space implies 
$\vb*{u}_{-\vb*{k}}=\vb*{u}_{\vb*{k}}^*$, and 
the incompressible condition is given by 
$\vb*{k}\cdot\vb*{u}_{\vb*{k}}=0$.  In the following, 
$a_j$ denotes the $j$-th component of the vector $\vb*{a}$ 
and the summation over repeated component indices is assumed.  

The NS equation in the wavevector space is given by 
\begin{equation}
\dv{t} \vb*{u}_{\vb*{k}}(t) = \vb*{M}_{\vb*{k}}\qty\big(\vb*{u}(t))
-\nu k^2\vb*{u}_{\vb*{k}}(t)+\vb*{f}_{\vb*{k}}(t), 
\label{eq:NSk}
\end{equation}
where the mass density of the fluid is unity, 
$\nu$ is the kinematic viscosity constant, 
$\vb*{f}(t)$ is the external forcing field, 
$\vb*{M}$ is a map from a vector field to a vector field 
whose component is given by 
\begin{align}
M_{\vb*{k},j}\qty(\vb*{u})&=-\frac{\ri}{2} 
\sum_{\vb*{p}}^{\Delta}\sum_{\vb*{q}}^{\Delta}
\delta_{\vb*{k}-\vb*{p}-\vb*{q}}^{\Delta}
\Biggl(k_m\qty(\delta_{jn}-\frac{k_jk_n}{k^2})
\nonumber\\
&\quad + k_n\qty(\delta_{jm}-\frac{k_jk_m}{k^2})\Biggr)
u_{\vb*{p},m} u_{\vb*{q},n}, 
\end{align}
$\sum_{\vb*{k}}^\Delta:=\sum_{\vb*{k}\in\mathcal{K}}\qty(\Delta k)^d$, 
$\delta^{\Delta}_{\vb*{k}}=(\Delta k)^{-d}$ for $\vb*{k}=\vb*{0}$ 
and $\delta^{\Delta}_{\vb*{k}}=0$ otherwise, and 
$\delta_{jm}$ is the Kronecker delta. 

The energy density per unit volume, or simply energy hereafter,  $E(\vb*{u})$ 
is given by 
\begin{equation}
E(\vb*{u})=\sum_{\vb*{k}}^\Delta E_{\vb*{k}}(\vb*{u}), \quad
E_{\vb*{k}}(\vb*{u}):=\frac{1}{2}\qty(\Delta k)^d |\vb*{u}_{\vb*{k}}|^2, 
\end{equation}
where $E_{\vb*{k}}(\vb*{u})$ is the energy for the wavevector mode $\vb*{k}$. 
Hereafter, let $\vb*{u}(t)$ denote the solution of (\ref{eq:NSk})
with $\nu=0$, $\vb*{f}(t)=\vb{0}$ and 
the initial condition $\vb*{u}$ at $t=0$.  
%The energy $E(\vb*{u})$ is conserved under the nonlinear interaction 
%represented by $\vb*{M}(\vb*{u})$, i.e., $E(\vb*{u}(t))=E(\vb*{u})$, and 
The energy flux $\Phi_{k}(\vb*{u})$ 
from the small-wavenumber region $\{\vb*{p}|p<k\}$ 
to the large-wavenumber region $\{\vb*{p}|p\ge k\}$ due to 
the interaction represented by $\vb*{M}$ is given by 
\begin{equation}
\Phi_{k}(\vb*{u})
%&:=-\dv{t}\sum_{\vb*{p}(p<k)}^\Delta \eval{E_{\vb*{p}}(\vb*{u}(t))}_{t=0}
%\nonumber\\&
:=-\qty(\Delta k)^d\sum_{\vb*{p}(p<k)}^\Delta 
\Re \qty(\vb*{M}_{\vb*{p}}(\vb*{u})\cdot\vb*{u}_{-\vb*{p}}). 
\label{eq:Phi}
\end{equation}

\section{Ensemble model}
\label{sec:ensemble_model}
%\paragraph{Ensemble model.}
An ensemble of states is specified by a probability density function 
$P(\vb*{u})$ which satisfies $P(\vb*{u})\ge 0$ and 
$\int \mathcal{D}\vb*{u} P(\vb*{u})=1$, where 
$\mathcal{D}\vb*{u}:=\prod_{\vb*{k}\in \mathcal{K}^+} \dd{\vb*{u}_{\vb*{k}}}, 
\dd\vb*{u}_{\vb*{k}}:=\prod_{j=1}^{d-1} \dd{\Re(u_{\vb*{k}}^{(j)})}\dd{\Im(u_{\vb*{k}}^{(j)})}$, 
$\mathcal{K}^+(\subset \mathcal{K})$ is a set of wavevectors such 
that either $\vb*{k}\in\mathcal{K}^+$ or $-\vb*{k}\in\mathcal{K}^+$ but 
not both for all $\vb*{k}\in\mathcal{K}$, and 
$u_{\vb*{k}}^{(j)}:=\vb*{u}_{\vb*{k}}\cdot \vb*{e}^{(j)}(\vb*{k})$ with 
$\vb*{e}^{(j)}(\vb*{k}) (j=1,\ldots,d-1)$ being a orthonormal-basis of 
the $d-1$-dimensional complex vector space perpendicular to $\vb*{k}$.  
The ensemble average of a function of the state $F(\vb*{u})$ is given by
$\langle F(\vb*{u})\rangle :=\int \mathcal{D}\vb*{u} P(\vb*{u}) F(\vb*{u})$.

We propose as an ensemble model of turbulence, the following 
probability density function, 
\begin{equation}
P_\epsilon(\vb*{u}):=C \prod_{n=0}^{N_t}\prod_{m=0}^{N_k}
\delta\qty\bigg(\Phi_{k_m}\qty\big(\vb*{u}(t_n))-\epsilon), 
\label{eq:Pfluxconst}
\end{equation}
where $\delta(x)$ is the Dirac delta function, $C$ is 
the constant for the normalization of probability, 
$\epsilon$ is a constant corresponding to the energy dissipation rate, 
$0<k_0<\ldots<k_{N_k}<k_{\max}$ and $0=t_0<t_1\ldots<t_{N_t}$. 
Formally, by taking limits $N_k,N_t\to \infty$ with 
$k_{\max},k_{N_k},t_{N_t}\to \infty$ and 
$\min_{m}(k_{m+1}-k_m), \min_{n}(t_{n+1}-t_n)\to 0$, 
one obtains a stationary ensemble model of 
states with the constant energy flux, $\Phi_k=\epsilon$
for $k\ge k_0$.  
%% An alternative expression of $P_\epsilon(\vb*{u})$ can be obtained 
%% by using auxiliary variables $\lambda_{k_m,t_n}$ as
%% \begin{align}
%% P_\epsilon(\vb*{u})&=C'\qty(\prod_{n=0}^{N_t}\prod_{m=0}^{N_k} 
%% \int_{-\infty}^\infty d\lambda_{k_m,t_n}) 
%% \nonumber\\
%% &\quad\times\exp \qty(\sum_{m=0}^{N_t}\sum_{n=0}^{N_k} \ri\lambda_{k_m,t_n}
%% \qty\bigg(\Phi_{k_m}\qty\big(\vb*{u}(t_n))-\epsilon)), 
%% \label{eq:Pfluxconstexp}
%% \end{align}
%% where $C'$ is the normalization constant. 

In the ensemble model $P_\epsilon(\vb*{u})$, 
the states are subject to the conditions 
$\Phi_{k_m}\qty\big(\vb*{u}(t_n))=\epsilon$ and 
the probability is distributed equally to the possible 
states in the sense that there is no other constraint.  
The model is similar to the microcanonical ensemble 
in which the states are subject to the condition that 
the energy is equal to a specific value.  
Behind the construction of the present ensemble model 
underlies the concept of typicality. 
The typicality implies that 
typical states (i.e., almost all states) $\vb*{u}$ in the ensemble 
already possess some properties of the ensemble average, i.e., 
$F(\vb*{u})\approx\langle F(\vb*{u})\rangle$ for the functions 
$F(\vb*{u})$ of interest.  
%Combined with the concept of typicality, 
It is supposed in the present ensemble model that 
each of the states of constant flux such that 
$\Phi_{k}\qty\big(\vb*{u}(t))=\epsilon$ for $k$ in the inertial range 
and $t$ in the time interval under consideration typically possesses 
a considerable part of the characteristics of turbulence.  
Note that 
a quasi-constant flux $\Phi_k\qty\big(\vb*{u}(t))\approx\epsilon$ is 
observed in many direct numerical simulations of the NS turbulence 
in the periodic boundary box, although the inertial range is limited. 
(See, e.g., Ref.\cite{Ishiharaetal2016}.)
The fact suggests that the constant flux in the inertial range 
is one of the essential characteristics of fully developed turbulence. 
The present ensemble model $P_\epsilon\qty\big(\vb*{u})$ would be appropriate 
if a considerable part of the other characteristics of turbulence 
can be derived from the property of constant flux. 

In spite of $P_\epsilon(\vb*{u})$ being a probability density function 
on the state space, the trajectory $\vb*{u}(t)$ is explicitly involved 
in the expressions of Eq.~(\ref{eq:Pfluxconst}). 
%and (\ref{eq:Pfluxconstexp}).  
For the sake of simplicity, let us replace $\Phi_{k}(\vb*{u}(t))$ in 
(\ref{eq:Pfluxconst}) by its $N_t$-th degree Taylor polynomial in $t$. 
Then, we may rewrite (\ref{eq:Pfluxconst}) as 
\begin{equation}
P_\epsilon^{(N_t)}(\vb*{u})=C'\prod_{m=0}^{N_k}\prod_{n=0}^{N_t}
\delta\qty(\Phi_{k_m}^{(n)}(\vb*{u})-\epsilon\delta_{n0}), 
\label{eq:PfluxconstNt}
\end{equation}
where
\begin{equation}
\Phi_{k}^{(n)}(\vb*{u}):=\eval{\dv[n]{t} \Phi_{k}\qty\big(\vb*{u}(t))}_{t=0}, 
\label{eq:Phikn}
\end{equation}
$C'$ is a normalizing constant, and 
we now write $N_t$ explicitly in the superscript for this approximation.  
The expression (\ref{eq:PfluxconstNt}) solely contains 
the instantaneous $\vb*{u}$.  The limit $N_t\to \infty$ should be taken 
in order that $P_\epsilon^{(N_t)}(\vb*{u})$ is stationary.  

Although the model of the ensemble is explicitly given in Eq.~(\ref{eq:Pfluxconst}) or (\ref{eq:PfluxconstNt}), there are some problems regarding the appropriateness of the model.  The existence of normalizing constants such that $C,C'>0$ for fixed $N_k$ and $N_t$ is not clear.  The suitable way of taking the limit $N_k, N_t \to \infty$ should be also discussed.  

Even if the problems of the appropriateness are solved or avoided in some way, computation of the ensemble average of quantities such as $E_{\vb*{k}}(\vb*{u})$ are difficult for $P_\epsilon(\vb*{u})$ or $P_\epsilon^{(N_t)}(\vb*{u})$ even with $N_t=0$.  This is because that $\Phi_k(\vb*{u})$ in Eq.~(\ref{eq:Pfluxconst}) or (\ref{eq:PfluxconstNt}) 
consists of third order terms in $\vb*{u}$ such as $\vb*{u}_{\vb*{k}}\vb*{u}_{\vb*{p}}\vb*{u}_{\vb*{q}} \delta_{\vb*{k}+\vb*{p}+\vb*{q}}^\Delta$ and that $\vb*{u}_{\vb*{k}}$ with different wavevectors $\vb*{k}$ are complexly coupled in $\Phi_k(\vb*{u})$.  It is desired to develop some analytical methods for the computation.  One candidate may be a method similar to the Martin-Siggia-Rose-Janssen-de Dominicis procedure \cite{MartinSiggiaRose1973, Janssen1976,deDominicis1976}.  The model $P_\epsilon(\vb*{u})$ can be expressed in a form that may be more familiar in the field theory by using auxiliary variables $\lambda_{k_m,t_n}$, as
\begin{align}
P_\epsilon(\vb*{u})&=C''\qty(\prod_{n=0}^{N_t}\prod_{m=0}^{N_k} 
\int_{-\infty}^\infty d\lambda_{k_m,t_n}) 
\nonumber\\
&\quad\times\exp \qty(\sum_{m=0}^{N_t}\sum_{n=0}^{N_k} \ri\lambda_{k_m,t_n}
\qty\bigg(\Phi_{k_m}\qty\big(\vb*{u}(t_n))-\epsilon)), 
\label{eq:Pfluxconstexp}
\end{align}
where $C''$ is the normalizing constant.  One may also consult Ref.~\cite{Sverak2017} for the treatment of probability measures with constraints imposed in the form of the Dirac delta function.  However, we will not pursue such analytical methods further in this study.

\section{Numerical sampling}
\label{sec:numericalsampling}
%\paragraph{Numerical sampling.}
If typicality applies to the present ensemble model, 
some properties of turbulence should be possessed by 
a single typical state in the ensemble before taking the average.  
Here, we attempt a random sampling from the ensemble model 
by a Monte Carlo (MC) method.  

For a first trial, we treat the case with $d=2$ for saving 
the computational resource.  In the case of $d=2$, the enstrophy 
\begin{equation}
\Omega(\vb*{u}):=(\Delta k)^d \frac{1}{2}\sum_{\vb*{k}}^\Delta  |\omega_{\vb*{k}}|^2 
=\sum_{\vb*{k}}^\Delta k^2 E_{\vb*{k}}(\vb*{u}), 
\end{equation}
where $\omega_{\vb*{k}}:=\ri (k_1u_{\vb*{k},2}- k_2u_{\vb*{k},1})$ is the 
vorticity field, is an inviscid conserved quantity 
as well as the energy.  Here, we consider the enstrophy cascade range.  
The enstrophy flux $\Phi^\Omega_k(\vb*{u})$, its time derivatives 
${\Phi^\Omega}_k^{(n)}(\vb*{u})$  and the probability density 
function $P_\eta^{(N_t)}(\vb*{u})$ of the constant-enstrophy-flux ensemble model, 
where $\eta$ is a constant corresponding to the enstrophy dissipation rate, 
can be defined similarly as in the case of 
$\Phi_k(\vb*{u})$ in Eq.~(\ref{eq:Phi}), 
$\Phi_k^{(n)}$ in Eq.~(\ref{eq:Phikn}), and 
$P_\epsilon^{(N_t)}(\vb*{u})$ in Eq.~(\ref{eq:PfluxconstNt}), respectively.  
%% The enstrophy flux from the region 
%% $\{\vb*{p}|p < k\}$ to $\{\vb*{p}|p\ge k\}$ due to the interaction 
%% is given by 
%% \begin{equation}
%% \Phi^\Omega_k(\vb*{u})=-(\Delta k)^d\sum_{\vb*{p}(p<k)}
%% p^2 \Re \qty(M(\vb*{u})_{\vb*{p},j} u_{-\vb*{p},j}). 
%% \end{equation}
%% The ensemble model of the constant enstrophy flux states 
%% is given by the probability density function 
%% \begin{equation}
%% P_\eta^{(N_t)}(\vb*{u})=C'''\prod_{m=0}^{N_k}\prod_{n=0}^{k_T}
%% \delta\qty({\Phi^\Omega}_{k_m}^{(n)}(\vb*{u})-\eta\delta_{n0}),
%% \label{eq:PfluxconstensNt}
%% \end{equation}
%% where $\eta$ is a constant corresponding to the enstrophy dissipation rate. 

Let us define the error functions by
\begin{equation}
\Delta^{(n)}(\vb*{u}):=\frac{1}{N_k+1}
\sum_{m=0}^{N_k}\qty({\Phi^\Omega}_{k_m}^{(n)}(\vb*{u})-\eta\delta_{n0})^2, 
\end{equation}
for $n=0,\ldots,N_t$.  
The MC step associated with  $\vb*{k}(\in\mathcal{K}^+)$,  
which updates a given state $\vb*{u}$ to a new one, 
is given by the following substeps. 
(1) Let 
\begin{equation}
\vb*{u}'_{\vb*{k}}=\vb*{u}_{\vb*{k}}\exp\qty(r \ee^{\ri\theta}),\quad
\vb*{u}'_{-\vb*{k}}=(\vb*{u}'_{\vb*{k}})^*, 
\end{equation}
and $\vb*{u}'_{\vb*{p}}=\vb*{u}_{\vb*{p}}$ for $\vb*{p}\ne \pm\vb*{k}$, 
where $r$ is a fixed parameter satisfying $0<r<1$ 
and $\theta$ is a uniform random variable on $[0,2\pi)$. 
(2) Accept $\vb*{u}'$ as the new state of $\vb*{u}$ with the probability
\begin{align}
T(\vb*{u}', \vb*{u})&=
\qty(\prod_{n=0}^{N_t}
\min\qty(\ee^{-\alpha^{(n)}(\Delta^{(n)}(\vb*{u}')-\Delta^{(n)}(\vb*{u}))},1))
\nonumber\\
&\quad\times\min\qty(\frac{|\vb*{u}'_{\vb*{k}}|^2}{|\vb*{u}_{\vb*{k}}|^2},1)
\label{eq:Tuu'},
\end{align}
and keep $\vb*{u}$ unchanged otherwise, 
where $\alpha^{(n)}(n=0,\ldots,N_t)$ 
are parameters satisfying $0\le \alpha^{(n)} \le \infty$.  
Since the typical scale of $\vb*{u}_{\vb*{k}}$ is not known a priori, 
we set a uniform step amplitude $r$ in $(\ln \vb*{u}_{\vb*{k}})$-space. 
The transition probability $T(\vb*{u}', \vb*{u})$ is that of 
the Metropolis algorithm 
with a modification factor due to the nonuniform step in $\vb*{u}_{\vb*{k}}$-space. 
The stationary probability density function concerning 
the MC steps for all $\vb*{k}\in\mathcal{K}^+$ satisfies
\begin{equation}
P_{\eta,\mathrm{MC}}^{(N_t)}(\vb*{u})\propto 
\ee^{-\sum_{n=0}^{N_t}\alpha^{(n)} \Delta^{(n)}(\vb*{u})}, 
\end{equation}
and $P_{\eta,\mathrm{MC}}^{(N_t)}(\vb*{u})$ tends to $P_\eta^{(N_t)}(\vb*{u})$ 
in the limit $\alpha^{(n)}\to \infty$.  

In this study, we deal with the ensemble model $P_\eta^{(0)}(\vb*{u})$.  
The numerical settings are as follows. 
The number of grid points in the periodic domain $[0,2\pi]^2$ is $N^2$. 
A Fourier spectral method with a phase-shift is used for the computation 
of the nonlinear terms and the maximum wavenumber is $k_{\max}=\sqrt{2} N/3$. 
The initial state of $\vb*{u}$ is generated randomly with the constraint 
that the enstrophy is equally distributed to all wavenumber modes, i.e., 
$\vb*{u}_{\vb*{k}}=(\beta_0)^{-1/2}k^{-1}\exp(\ri\theta_{\vb*{k}})
\vb*{e}^{(1)}(\vb*{k})$ with $0<\beta_0<\infty$ and 
$\theta_{\vb*{k}}$ being uniform random variables on $[0,2\pi)$.  
An MC cycle is defined by the performance of the MC steps associated 
with $\vb*{k}$ for all $\vb*{k}\in\mathcal{K}^+$ in the order of increasing $k$.  
The values of the parameters are  
$N=512$, $N_k=239$, $k_m=m+1.5 (0\le m \le N_k)$, 
$\eta=1$, $r=0.0625$, $\alpha^{(0)}=\infty$, and $\beta_0=10^8$.  
Four sequences of MC cycles with different random seeds, SEQ0 to SEQ3,
are performed up to $c=800$ where $c$ is the number of MC cycles. 

\begin{figure}
\includegraphics[width=0.45\textwidth]{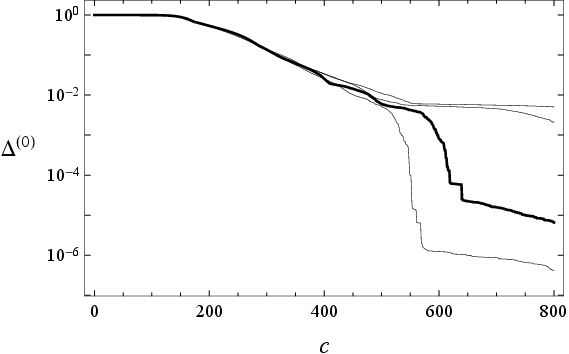}
\caption{The error function $\Delta^{(0)}(\vb*{u})$ of the realized states 
$\vb*{u}$ in the Monte Carlo cycles as a function of 
the number of cycles $c$.  
The thick line corresponds to the sequence SEQ0 and 
the thin lines to the other sequences, SEQ1 to SEQ3. 
}
\label{fig:fluxerr}
\end{figure}
\begin{figure}
\includegraphics[width=0.45\textwidth]{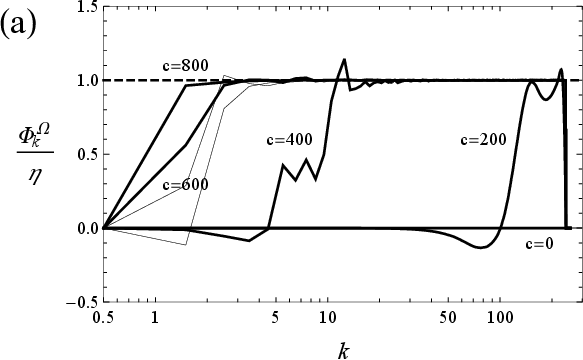}
\includegraphics[width=0.45\textwidth]{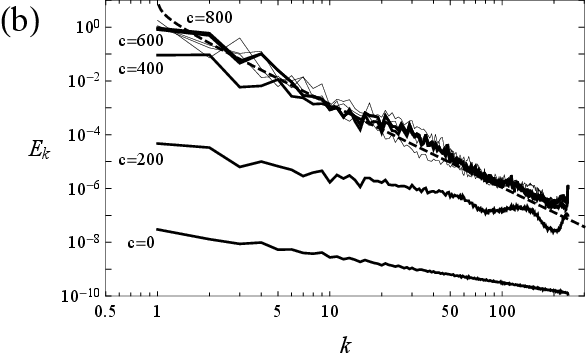}
\caption{(a) The enstrophy flux $\Phi^{\Omega}_k(\vb*{u})$ normalized 
by the parameter $\eta$ corresponding to the enstrophy dissipation rate 
and (b) the energy spectrum $E_k(\vb*{u})$, of the realized states $\vb*{u}$ 
in the sequence SEQ0 of the Monte Carlo cycles where $c$ is the number of 
the cycles (thick lines). The thin lines 
show results at $c=800$ from the other sequences, SEQ1 to SEQ3. 
The dashed line in (b) shows 
$E_k=C_K \eta^{2/3}k^{-3}(\ln(k/k_\mathrm{b}))^{-1/3}$ with $C_K=1.81$(LRA) and $k_\mathrm{b}=1$.
}
\label{fig:fluxespe}
\end{figure}
We can confirm that the error function $\Delta^{(0)}(\vb*{u})$ decreases 
with the increase of $c$ in Fig.~\ref{fig:fluxerr}. 
The enstrophy flux $\Phi^\Omega_k(\vb*{u})$ of the states $\vb*{u}$ 
obtained by the MC method are also shown in Fig.~\ref{fig:fluxespe}(a).  
It is observed that the wavenumber region such that 
$\Phi^\Omega_k(\vb*{u})\approx\eta$ expands from large to 
small wavenumbers with the increase of $c$.  
The relative discrepancy from the constant flux 
$|\Phi^\Omega_k(\vb*{u})-\eta|/\eta$ is smaller than $0.025$ 
in the wavenumber range $5.5 \le k \le 240.5$ at $c=800$ for all the sequences.  
%We may consider that the states $\vb*{u}$ at $c=800$ are 
%those sampled out of the ensemble of states with 
%a constant enstrophy flux at the level of $N_t=0$ 
%in the wavenumber range and the precision noted above.  

%It is of interest that to what extent these sampled states possess 
%the properties of turbulence.  
The energy spectrum of the state $\vb*{u}$ defined by 
\begin{equation}
E_k(\vb*{u}):=(\Delta k)^{-1}
\sum_{\substack{\vb*{p}\\(k-\Delta k/2 \le p<k+\Delta k/2)}}^{\Delta}
E_{\vb*{p}}(\vb*{u})
\end{equation}
is given for the states obtained by the MC method in 
Fig.\ref{fig:fluxespe}(b).  It is found that 
$E_{k}(\vb*{u})$ tends to converge with the increase 
of $c$ and that 
$E_k(\vb*{u})$ at $c=800$ in all the sequences are close to 
the energy spectrum in the enstrophy cascade range, 
$E_k=C_K \eta^{2/3}k^{-3}(\ln(k/k_\mathrm{b}))^{-1/3}$ with $C_K=1.81$,  
estimated in the LRA\cite{Kaneda1987,Kaneda2007}, 
where we put the bottom wavenumber of the inertial range 
as $k_\mathrm{b}=1$.  It is known that $E_k$ in the LRA 
is in good agreement with the results from the numerical 
simulations\cite{IshiharaKaneda2001}. See also Appendix. 
The ratio of the energy spectra by the MC method 
$E_k^{(\mathrm{MC})}(\vb*{u})$ at $c=800$ to 
that of the LRA $E_k^{(\mathrm{LRA})}$ are confined 
in the range $0.4<E_k^{(\mathrm{MC})}(\vb*{u})/E_k^{(\mathrm{LRA})}<7.5$ 
for the wavenumber range $3\le k \le 239$.  A general tendency is 
that the ratio increases near the edge $k=k_{\max}$.  

\begin{figure}
\includegraphics[width=0.23\textwidth]{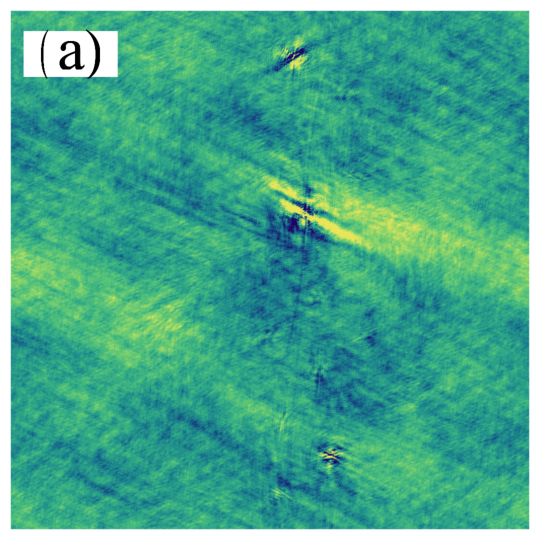}
\includegraphics[width=0.23\textwidth]{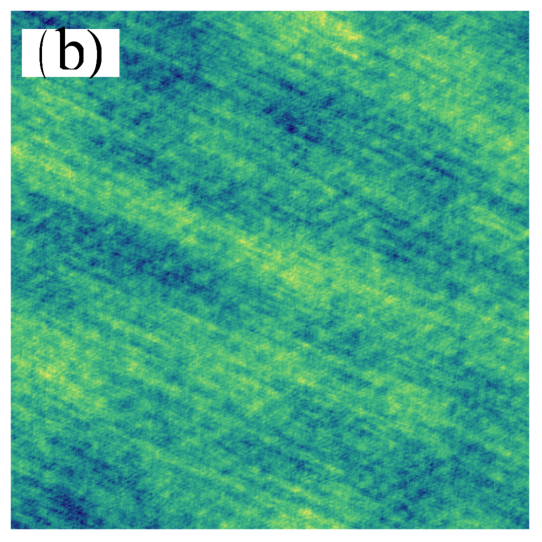}
\caption{(a) Vorticity field in the real space of the state $\vb*{u}$ 
obtained by the Monte Carlo method at $c=800$ of SEQ0.  
(b) Phase randomized vorticity field with the same 
energy $E_{\vb*{k}}$ for each wavevector $\vb*{k}$ as (a).  
Bright(dark) regions correspond to positive(negative)-vorticity regions.}
\label{fig:vorx}
\end{figure}
The vorticity field in the real space of a state obtained at 
$c=800$ is given in Fig.~\ref{fig:vorx}(a) together with 
the phase-randomized field with the same energy $E_{\vb*{k}}$ for 
each wavevector $\vb*{k}$ as the original field in Fig.~\ref{fig:vorx}(b).  
One can observe some organized structures with intense vorticity 
in the state obtained from the MC method.  
The maximum absolute value of the vorticity is $5$ times larger than 
that of the phase-randomized field.  Since the structures are 
absent in the phase-randomized field, it is suggested that 
the emergence of the structures is due to the constraint of 
constant enstrophy flux.  Apparent anisotropy observed in both fields may 
be a sign of prominent amplitudes of some specific wavevector modes. 

\section{Discussions}
\label{sec:discussions}
%\paragraph{Discussions.}
We proposed an ensemble model of turbulence $P_\epsilon(\vb*{u})$ which can be constructed explicitly by Eq.(\ref{eq:Pfluxconst}).  Since the probability measure is uniformly distributed for the states of constant flux with the value of flux $\epsilon$, the ensemble model $P_\epsilon(\vb*{u})$ maximizes the entropy $S(P)=-\int \mathcal{D}\vb*{u} P(\vb*{u}) \ln P(\vb*{u})$ within the constraint that $P(\vb*{u})\ne 0$ only for $\vb*{u}$ being one of the states of constant flux. Here we recall the entropy method (EM) for turbulence proposed by Edwards and McComb\cite{EdwardsMcComb1969}.  The model probability density function $P(\vb*{u})$ in the method is parameterized by $\phi_l$ and $\eta_l$, where $l$ runs over all possible components of the velocity field modes $\vb*{u}_{\vb*{k},j}$.  (See Ref.\cite{inLeslie1973_7} for the notations.)  The parameters $\phi_l$ and $\eta_l$ are associated with the intensity and damping rate, respectively, of the corresponding mode $\vb*{u}_{\vb*{k},j}$.  The model $P(\vb*{u})$ in EM is obtained by a perturbation from a multivariate normal distribution such that $P(\vb*{u})>0$ for all $\vb*{u}$ provided that $\phi_l\ne 0$ for all $l$.  The model $P_\epsilon(\vb*{u})$ in Eq.(\ref{eq:Pfluxconst}) is quite different from the multivariate normal distribution since $P(\vb*{u})=0$ for all $\vb*{u}$ unless $\vb*{u}$ is one of the states of constant flux.  Therefore, it is likely that $P_\epsilon(\vb*{u})$ is not included in the class of probability density function $P(\vb*{u})$ considered in EM.  We do not attempt the proof here.  It is argued in Ref.~\cite{Qian1996} that the maximum entropy state under the constraint of constant flux does not exist in the formalism of EM.  It is not obvious that the argument can be extended to the present model, $P_\epsilon(\vb*{u})$ in Eq.(\ref{eq:Pfluxconst}).  In the context of the present study, the existence of $P_\epsilon(\vb*{u})$ that maximizes the entropy under the constraint would be related to the existence of the normalizing constant $C>0$ that is referred to in Sec.~\ref{sec:ensemble_model}.  The related analysis will be left for a future study. 
%The model is essentially determined by the form of the nonlinear 
%interaction $\vb*{M}(\vb*{u})$ and the value of flux $\epsilon$. 

One way for the validation of the present ensemble model $P_\epsilon(\vb*{u})$ in Eq.(\ref{eq:Pfluxconst}) is to compute the ensemble averages of some quantities and then compare them with known results in the turbulence statistics.  However, the analytical methods for the computation are yet to be developed even for the approximate expression $P_\epsilon^{(0)}(\vb*{u})$ in Eq.(\ref{eq:PfluxconstNt}).  The way we took in this study is the numerical sampling from the ensemble, which is rather accessible.  

It should be noted that the numerical sampling in this study is at 
a beginning stage and that the possibility of some bias in the sampling 
is not excluded.  
We cannot conclude whether the anisotropy in the sampled vorticity field 
is a genuine feature of the constant flux states or an artifact of 
the sampling at the current stage.  
In the present analysis, the amplitude of the initial states is chosen to be small so that the energy spectra $E_k(\vb*{u})$ of 
the constant flux states are approached from below.  
The MC sequences with an initially large amplitude are not satisfactory 
so far.  Shortage of the analysis aside, it is remarkable that 
the energy spectra $E_{\vb*{k}}(\vb*{u})$ of the states sampled out from 
the ensemble $P_{\eta}^{(0)}(\vb*{u})$ of the constant enstrophy flux states 
in two-dimensional turbulence are consistent with the form 
$E_{\vb*{k}}=C_K\eta^{2/3}k^{-3}(\ln(k/k_\mathrm{b}))^{-1/3}$ that is obtained in 
the closure theories and verified in the numerical simulations of the NS equation.  
A positive prospect is that the ensemble model 
$P_\eta^{(N_t)}(\vb*{u})$ and the associated random sampling 
can be useful to analyze some turbulence statistics 
at relatively low approximation levels such as $N_t=0$ or $1$. 
Although the constraint of the constant flux ${\Phi^\Omega}_k=\eta$ 
in wavenumber space, i.e., the ensemble model $P_{\eta}^{(0)}(\vb*{u})$, 
yields some spatial structures of the vorticity field, 
they do not resemble those in the numerical simulations of the two dimensional 
turbulence. See Appendix or e.g., Ref. \cite{BoffettaEcke2012}.  
It would be of interest to investigate how the further constraints on 
the time derivatives of the flux ${\Phi^\Omega}_k^{(N_t)} (N_t \ge 1)$, 
i.e., the applications of the ensemble models $P_\eta^{(N_t)}(N_t \ge 1)$, affect the structures of the vorticity field as well as the spectrum and higher-order moments.  
 \\

\section*{Acknowledgement}
The author is grateful to Yukio Kaneda for valuable discussions 
including the one that inspired the numerical sampling of the present work. 
The author also thanks Yasuhiro Tokura for valuable discussions.  
This research was supported by Multidisciplinary Cooperative Research Program 
in Center for Computational Sciences, University of Tsukuba.  

\appendix*

\section{Numerical simulations of two-dimensional Navier-Stokes equation}
We performed numerical simulations of two-dimensional turbulence with random forcing and hyperviscosity in a periodic boundary box.  Basically, we followed the setting of the simulations in Ref.~\cite{IshiharaKaneda2001}.  The governing equation of the simulations is given by 
\begin{equation}
\pdv{t}\omega_{\vb*{k}}=J_{\vb*{k}}+ d_{\vb*{k}} + f_{\vb*{k}}, 
\end{equation}
where $J_{\vb*{k}}$ is the Fourier transform of the Jacobian given by 
$J(\psi,\omega)=\partial_1 \psi \partial_2\omega -\partial_2\psi\partial_1\omega$ in the physical space, 
$\psi_{\vb*{k}}$ is the stream function related to the velocity field by 
$\vb*{u}_{\vb*{k}}=(\ri k_2 \psi_{\vb*{k}}, -\ri k_1 \psi_{\vb*{k}})$ and to the vorticity field by $\omega_{\vb*{k}}= k^2\psi_{\vb*{k}}$, $d_{\vb*{k}}$ is the dissipation term and $f_{\vb*{k}}$ is the forcing term.  The nonlinear term $J_{\vb*{k}}$ is computed in the same way as the numerical sampling in Sec.~\ref{sec:numericalsampling}.  The dissipation term $d_{\vb*{k}}$ is given by 
\begin{equation}
d_{\vb*{k}}=-\nu \qty(2\Omega)^{\frac{1}{2}}\qty(\frac{k}{k_{\max}})^6 
\omega_{\vb*{k}} -\gamma \chi_{(0,k_\gamma)}(k)\omega_{\vb*{k}},
\end{equation}
where $\nu$ is the coefficient for the hyperviscosity, $\Omega$ is the enstrophy calculated at every time step, $\gamma$ is the coefficient of the drag applied in the wavenumber range $0<k<k_\gamma$, and $\chi_A(k)$ is a function such that $\chi_A(k)=1$ for $k\in A$ and $\chi_A(k)=0$ otherwise.  The drag in the small wavenumber range prevents the accumulation of the energy that cascades inversely into the range.  
The forcing term $f_{\vb*{k}}$ is given by 
\begin{equation}
f_{\vb*{k}}=\chi_{[k_{\mathrm{fmin}},k_{\mathrm{fmax}})}(k) 
\qty(2 \eta_{\mathrm{f}})^{\frac{1}{2}}N_{\mathrm{f}}^{-\frac{1}{2}}\qty(\Delta t)^{-\frac{1}{2}}
\qty(\Delta k)^{-2}
\ee^{\ri\varphi_{\vb*{k}}}, 
\end{equation}
where $\eta_{\mathrm{f}}$ is the average enstrophy injection rate by the forcing, $\Delta t$ is the time increment in the simulations, $N_{\mathrm{f}}$ is the number of wavevectors $\vb*{k}$ satisfying $k_{\mathrm{fmin}}\le k <k_{\mathrm{fmax}}$, $\varphi_{\vb*{k}} (\vb*{k}\in \mathcal{K}')$ is a uniform random variable on $[0,2\pi)$ generated at every time step and $f_{-\vb*{k}}=f_{\vb*{k}}^*$.  

The values of the parameters in the simulation are as follows.  The number of grid points along one coordinate direction is $N=4096$, the length of the sides of the domain is $L=2\pi$ which implies $\Delta k=1$, $k_{\max}=(\sqrt{2} N/3) \Delta k =1931$, $\Delta t=0.16\times 10^{-3}$, $\nu=1.0$, $\gamma=1.0$, $k_\gamma=2.5$, $\eta_{\mathrm{f}}=1.0$, $k_{\mathrm{fmin}}=4.5$, and $k_{\mathrm{fmax}}=7.5$.  The initial state $\omega_{\vb*{k}}(t=0)$ was generated under the conditions $|\omega_{\vb*{k}}(t=0)|\propto k^2 \exp(-3 k^2/2 k_a^2)$, $k_a=8$ and the enstrophy $\Omega(t=0)=1$.  The phase of $\omega_{\vb*{k}}(t=0)$ was determined randomly.  

\begin{figure}
\includegraphics[width=0.45\textwidth]{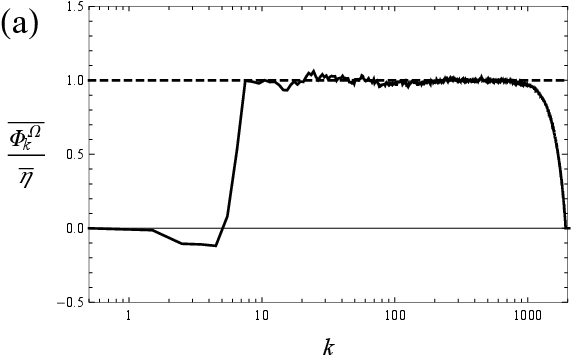}
\includegraphics[width=0.45\textwidth]{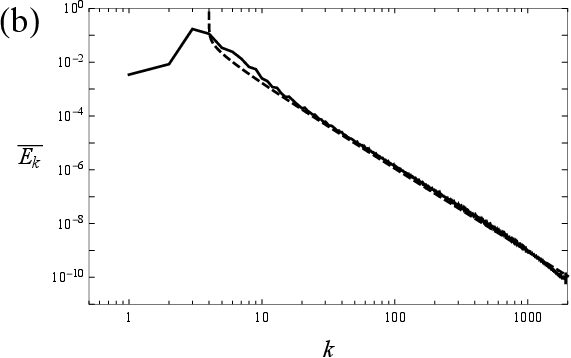}
\caption{(a) The time-averaged enstrophy flux $\overline{\Phi_k^\Omega}$ normalized by the time-averaged enstrophy dissipation rate $\overline{\eta}$ and (b) the time-averaged energy spectrum $\overline{E_k}$ in the numerical simulation.  The dashed line in (b) shows $E_k=C_K\eta^{2/3}k^{-3}\qty(\ln(k/k_{\mathrm{b}}))^{-1/3}$ with $C_K=1.81$(LRA) and $k_{\mathrm{b}}=4.0$.}
\label{fig:dnsflsp}
\end{figure}

The enstrophy dissipation rate $\eta(t)$ become quasi-stationary for $t \gtrsim 64$. Hereafter, $\overline{x}$ denotes the time average of $x$ in the time interval $64 \le t \le 128$.  It is observed that $\overline\eta = 0.867$ and the normalized standard deviation $\qty(\overline{(\eta -\overline\eta)^2})^{1/2}/\overline\eta$ is $0.063$.  The time-averaged enstrophy flux $\overline{\Phi_k^\Omega}$ is given in Fig.~\ref{fig:dnsflsp} (a).  The normalized deviation of the time-averaged enstrophy flux from the time-averaged enstrophy dissipation rate $|\overline{\Phi_k^\Omega} -\overline{\eta}|/\overline{\eta}$ is smaller than $0.05$ in the wavenumber range $26 \le k \le 1100$.  Here, we consider the wavenumber range as the enstrophy cascade range.  The normalized standard deviation $\qty(\overline{(\Phi_k^\Omega -\overline{\Phi_k^\Omega})^2})^{1/2}/\overline{\Phi_k^\Omega}$ is smaller than $0.25$ in the enstrophy cascade range.  

The time-averaged energy spectrum $\overline{E_k}$ is given in Fig.~\ref{fig:dnsflsp}(b). The normalized standard deviation $\qty(\overline{(E_k-\overline{E_k})^2})^{1/2}/\overline{E_k}$ is smaller than $0.15$ in the enstrophy cascade range.  The slope of $\overline{E_k}$ is slightly steeper than $k^{-3}$ and the functional form of $E_k(C_K,k_{\mathrm{b}})=C_K\overline{\eta}^{2/3}k^{-3}\qty(\ln(k/k_{\mathrm{b}}))^{-1/3}$ can be fitted to $\overline{E_k}$ with $C_K=2.20$ and $k_{\mathrm{b}}=4.0$ 
by minimizing the function 
$\sum_{k} \qty(\ln\qty(E_k(C_K,k_{\mathrm{b}})/\overline{E_k}))^2 \qty(\Delta k/k)$ 
where the summation is taken over $k$ in the enstrophy cascade range.  
The estimate of $C_K^{\mathrm{LRA}}=1.81$ in the LRA is in good agreement with that of the simulation in the sense that $|C_K^{\mathrm{LRA}}-C_K|/C_K <0.2$.  The energy spectrum $E_k^{\mathrm{LRA}}$ for the enstrophy cascade range estimated in LRA  with $k_{\mathrm{b}}=4.0$ is also plotted with a dashed line in Fig.~\ref{fig:dnsflsp}(b).  Note that $\overline{E_k}\propto k^{-3}$ without logarithmic correction may be observed when there is a sufficient amount of energy in the wavenumber range $k<k_{\mathrm{b}}$.  See Ref.~\cite{KanedaIshihara2001} for the detail.  Since the energy outside the inertial range is not considered in the numerical sampling in Sec.~\ref{sec:numericalsampling}, the present setting of the numerical simulation of the NS equation and the spectrum with the logarithmic correction may be appropriate for the comparison.  

%The energy spectrum $\overline{E_k}$ is in good agreement with 
%$E_k^{\mathrm{LRA}}$ in the sense that the ratio falls in 
%the range $0.95 < \overline{E_k}/E_k^{\mathrm{LRA}} < 1.3$ for 
%the enstrophy cascade range.  
%Here, we chose the bottom wavenumber $k_{\mathrm{b}}$ of the inertial range 
%to be the top wavenumber of the forcing range.  
%The energy spectrum $E_k^{\mathrm{LRA}}$ in a large part of 
%the enstrophy cascade range is somewhat insensitive to 
%the choice of $k_{\mathrm{b}}$.  For example, the choice $k_{\mathrm{b}}=26$ 
%increases $E_k^{\mathrm{LRA}}$ at $k=300$ only by the factor of $1.15$. 

\begin{figure}
\includegraphics[width=0.23\textwidth]{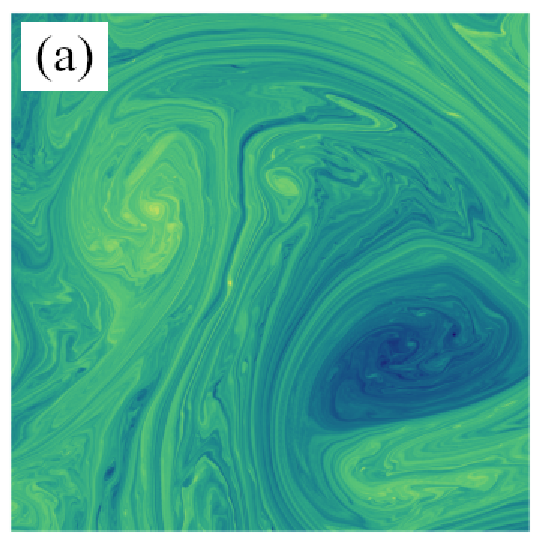}
\includegraphics[width=0.23\textwidth]{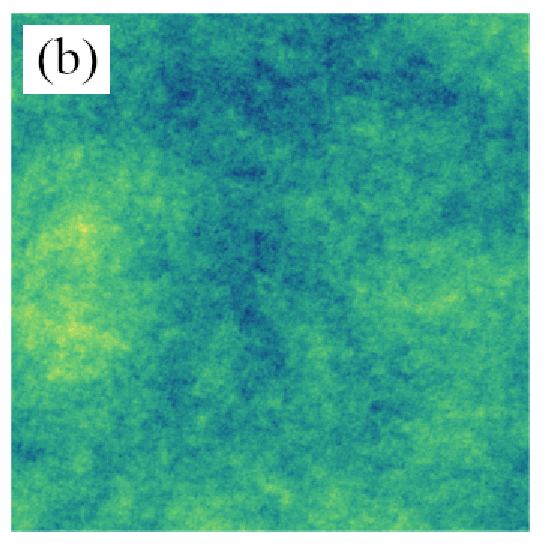}
\caption{
(a) Vorticity field in a subdomain with sides $\pi/2 \times \pi/2$ of 
the real space at $t=128$ of the simulation.  
(b) Phase randomized vorticity field with the same energy $E_{\vb*{k}}$ 
for each wavevector $\vb*{k}$ as (a).  
Bright(dark) regions correspond to positive(negative)-vorticity regions. 
}
\label{fig:dnsvorx}
\end{figure}
The vorticity field in the real space is given in 
Fig.~\ref{fig:dnsvorx} (a) for the simulated field at $t=128$.  
A subdomain with sides $\pi/2 \times \pi/2$ is displayed as 
a representative.  One can observe stretched and folded structures 
of iso-vorticity regions.  The structures disappear in 
the phase randomized vorticity fields with the same energy $E_{\vb*{k}}$ 
for each wavevector $\vb*{k}$ as shown in Fig.~\ref{fig:dnsvorx} (b).  


\begin{thebibliography}{29}%
\makeatletter
\providecommand \@ifxundefined [1]{%
 \@ifx{#1\undefined}
}%
\providecommand \@ifnum [1]{%
 \ifnum #1\expandafter \@firstoftwo
 \else \expandafter \@secondoftwo
 \fi
}%
\providecommand \@ifx [1]{%
 \ifx #1\expandafter \@firstoftwo
 \else \expandafter \@secondoftwo
 \fi
}%
\providecommand \natexlab [1]{#1}%
\providecommand \enquote  [1]{``#1''}%
\providecommand \bibnamefont  [1]{#1}%
\providecommand \bibfnamefont [1]{#1}%
\providecommand \citenamefont [1]{#1}%
\providecommand \href@noop [0]{\@secondoftwo}%
\providecommand \href [0]{\begingroup \@sanitize@url \@href}%
\providecommand \@href[1]{\@@startlink{#1}\@@href}%
\providecommand \@@href[1]{\endgroup#1\@@endlink}%
\providecommand \@sanitize@url [0]{\catcode `\\12\catcode `\$12\catcode
  `\&12\catcode `\#12\catcode `\^12\catcode `\_12\catcode `\%12\relax}%
\providecommand \@@startlink[1]{}%
\providecommand \@@endlink[0]{}%
\providecommand \url  [0]{\begingroup\@sanitize@url \@url }%
\providecommand \@url [1]{\endgroup\@href {#1}{\urlprefix }}%
\providecommand \urlprefix  [0]{URL }%
\providecommand \Eprint [0]{\href }%
\providecommand \doibase [0]{https://doi.org/}%
\providecommand \selectlanguage [0]{\@gobble}%
\providecommand \bibinfo  [0]{\@secondoftwo}%
\providecommand \bibfield  [0]{\@secondoftwo}%
\providecommand \translation [1]{[#1]}%
\providecommand \BibitemOpen [0]{}%
\providecommand \bibitemStop [0]{}%
\providecommand \bibitemNoStop [0]{.\EOS\space}%
\providecommand \EOS [0]{\spacefactor3000\relax}%
\providecommand \BibitemShut  [1]{\csname bibitem#1\endcsname}%
\let\auto@bib@innerbib\@empty
%</preamble>
\bibitem [{\citenamefont {Gibbs}(1902)}]{Gibbs1902}%
  \BibitemOpen
  \bibfield  {author} {\bibinfo {author} {\bibfnamefont {J.~W.}\ \bibnamefont
  {Gibbs}},\ }\href@noop {} {\emph {\bibinfo {title} {ElementaryoPrinciples in
  Statistical Mechanics: Developed with Especial Reference to the Rational
  Foundation of Thermodyncamics}}}\ (\bibinfo  {publisher} {Charles Scribner's
  Sons},\ \bibinfo {year} {1902})\BibitemShut {NoStop}%
\bibitem [{\citenamefont {Inaba}(2021)}]{Inaba2021}%
  \BibitemOpen
  \bibfield  {author} {\bibinfo {author} {\bibfnamefont {H.}~\bibnamefont
  {Inaba}},\ }\href@noop {} {\emph {\bibinfo {title} {The Making of Statistical
  Mechanics}}}\ (\bibinfo  {publisher} {The University of Nagoya Press},\
  \bibinfo {year} {2021})\ \bibinfo {note} {(in Japanese)}\BibitemShut
  {NoStop}%
\bibitem [{\citenamefont {Tasaki}(2016)}]{Tasaki2016}%
  \BibitemOpen
  \bibfield  {author} {\bibinfo {author} {\bibfnamefont {H.}~\bibnamefont
  {Tasaki}},\ }\href@noop {} {\bibfield  {journal} {\bibinfo  {journal}
  {J. Stat. Phys.}\ }\textbf {\bibinfo {volume} {163}},\
  \bibinfo {pages} {937} (\bibinfo {year} {2016})}\BibitemShut {NoStop}%
\bibitem [{\citenamefont {Goldstein}\ \emph {et~al.}(2020)\citenamefont
  {Goldstein}, \citenamefont {Lebowitz}, \citenamefont {Tumulka},\ and\
  \citenamefont {Zangh\`i}}]{GoldsteinLebowitzTumulkaZanghi2020}%
  \BibitemOpen
  \bibfield  {author} {\bibinfo {author} {\bibfnamefont {S.}~\bibnamefont
  {Goldstein}}, \bibinfo {author} {\bibfnamefont {J.~L.}\ \bibnamefont
  {Lebowitz}}, \bibinfo {author} {\bibfnamefont {R.}~\bibnamefont {Tumulka}},\
  and\ \bibinfo {author} {\bibfnamefont {N.}~\bibnamefont {Zangh\`i}},\ }in\
  \href@noop {} {\emph {\bibinfo {booktitle} {Statistical Mechanics and
  Scientific Explanation, Determinism, Indeterminism and Laws of Nature}}},\
  \bibinfo {editor} {edited by\ \bibinfo {editor} {\bibfnamefont
  {V.}~\bibnamefont {Allori}}}\ (\bibinfo  {publisher} {World Scientific},\
  \bibinfo {year} {2020})\ pp.\ \bibinfo {pages} {519--581}\BibitemShut
  {NoStop}%
\bibitem [{\citenamefont {Yoshida}(2020)}]{Yoshida2020}%
  \BibitemOpen
  \bibfield  {author} {\bibinfo {author} {\bibfnamefont {K.}~\bibnamefont
  {Yoshida}},\ }\href@noop {} {\bibfield  {journal} {\bibinfo  {journal} {Phys.
  Rev. A}\ }\textbf {\bibinfo {volume} {101}},\ \bibinfo {pages} {032110}
  (\bibinfo {year} {2020})}\BibitemShut {NoStop}%
\bibitem [{\citenamefont {Bedrossian}\ \emph {et~al.}(2022)\citenamefont
  {Bedrossian}, \citenamefont {Blumenthal},\ and\ \citenamefont
  {Punshon-Smith}}]{Bedrossian2022batchelor}%
  \BibitemOpen
  \bibfield  {author} {\bibinfo {author} {\bibfnamefont {J.}~\bibnamefont
  {Bedrossian}}, \bibinfo {author} {\bibfnamefont {A.}~\bibnamefont
  {Blumenthal}},\ and\ \bibinfo {author} {\bibfnamefont {S.}~\bibnamefont
  {Punshon-Smith}},\ }\href@noop {} {\bibfield  {journal} {\bibinfo  {journal}
  {Comm. Pure Appl. Math.}\ } (\bibinfo {year}
  {2022})},\ \bibinfo {note} {doi:
  https://doi.org/10.1002/cpa.22022}\BibitemShut {NoStop}%
\bibitem [{\citenamefont {Kawahara}\ and\ \citenamefont
  {Kida}(2001)}]{KawaharaKida2001}%
  \BibitemOpen
  \bibfield  {author} {\bibinfo {author} {\bibfnamefont {G.}~\bibnamefont
  {Kawahara}}\ and\ \bibinfo {author} {\bibfnamefont {S.}~\bibnamefont
  {Kida}},\ }\href@noop {} {\bibfield  {journal} {\bibinfo  {journal} {J. 
  Fluid Mech.}\ }\textbf {\bibinfo {volume} {449}},\ \bibinfo {pages}
  {291} (\bibinfo {year} {2001})}\BibitemShut {NoStop}%
\bibitem [{\citenamefont {Martin}\ \emph {et~al.}(1973)\citenamefont {Martin},
  \citenamefont {Siggia},\ and\ \citenamefont {Rose}}]{MartinSiggiaRose1973}%
  \BibitemOpen
  \bibfield  {author} {\bibinfo {author} {\bibfnamefont {P.}~\bibnamefont
  {Martin}}, \bibinfo {author} {\bibfnamefont {E.}~\bibnamefont {Siggia}},\
  and\ \bibinfo {author} {\bibfnamefont {H.}~\bibnamefont {Rose}},\ }\href@noop
  {} {\bibfield  {journal} {\bibinfo  {journal} {Phys. Rev. A}\ }\textbf
  {\bibinfo {volume} {8}},\ \bibinfo {pages} {423} (\bibinfo {year}
  {1973})}\BibitemShut {NoStop}%
\bibitem [{\citenamefont {Janssen}(1976)}]{Janssen1976}%
  \BibitemOpen
  \bibfield  {author} {\bibinfo {author} {\bibfnamefont {H.-K.}\ \bibnamefont
  {Janssen}},\ }\href@noop {} {\bibfield  {journal} {\bibinfo  {journal}
  {Z. Phys. B}\ }\textbf {\bibinfo {volume} {23}},\ \bibinfo
  {pages} {377} (\bibinfo {year} {1976})}\BibitemShut {NoStop}%
\bibitem [{\citenamefont {de~Dominicis}(1976)}]{deDominicis1976}%
  \BibitemOpen
  \bibfield  {author} {\bibinfo {author} {\bibfnamefont {C.}~\bibnamefont
  {de~Dominicis}},\ }\href@noop {} {\bibfield  {journal} {\bibinfo  {journal}
  {J. Phys. Colloques}\ }\textbf {\bibinfo {volume} {37}},\ \bibinfo
  {pages} {247} (\bibinfo {year} {1976})}\BibitemShut {NoStop}%
\bibitem [{\citenamefont {Canet}\ \emph {et~al.}(2016)\citenamefont {Canet},
  \citenamefont {Delamotte},\ and\ \citenamefont
  {Wschebor}}]{CanetDelamotteWschebor2016}%
  \BibitemOpen
  \bibfield  {author} {\bibinfo {author} {\bibfnamefont {L.}~\bibnamefont
  {Canet}}, \bibinfo {author} {\bibfnamefont {B.}~\bibnamefont {Delamotte}},\
  and\ \bibinfo {author} {\bibfnamefont {N.}~\bibnamefont {Wschebor}},\
  }\href@noop {} {\bibfield  {journal} {\bibinfo  {journal} {Phys. Rev. E}\
  }\textbf {\bibinfo {volume} {93}},\ \bibinfo {pages} {063101} (\bibinfo
  {year} {2016})}\BibitemShut {NoStop}%
\bibitem [{\citenamefont {Edwards}\ and\ \citenamefont
  {McComb}(1969)}]{EdwardsMcComb1969}%
  \BibitemOpen
  \bibfield  {author} {\bibinfo {author} {\bibfnamefont {S.}~\bibnamefont
  {Edwards}}\ and\ \bibinfo {author} {\bibfnamefont {W.}~\bibnamefont
  {McComb}},\ }\href@noop {} {\bibfield  {journal} {\bibinfo  {journal}
  {J. Phys. A: Gen. Phys.}\ }\textbf {\bibinfo {volume} {2}},\
  \bibinfo {pages} {157} (\bibinfo {year} {1969})}\BibitemShut {NoStop}%
\bibitem [{\citenamefont {Kraichnan}(1965)}]{Kraichnan1965}%
  \BibitemOpen
  \bibfield  {author} {\bibinfo {author} {\bibfnamefont {R.~H.}\ \bibnamefont
  {Kraichnan}},\ }\href@noop {} {\bibfield  {journal} {\bibinfo  {journal}
  {Phys. Fluids}\ }\textbf {\bibinfo {volume} {8}},\ \bibinfo {pages} {575}
  (\bibinfo {year} {1965})}\BibitemShut {NoStop}%
\bibitem [{\citenamefont {Kaneda}(1981)}]{Kaneda1981}%
  \BibitemOpen
  \bibfield  {author} {\bibinfo {author} {\bibfnamefont {Y.}~\bibnamefont
  {Kaneda}},\ }\href@noop {} {\bibfield  {journal} {\bibinfo  {journal} {J.
  Fluid Mech.}\ }\textbf {\bibinfo {volume} {107}},\ \bibinfo {pages} {131}
  (\bibinfo {year} {1981})}\BibitemShut {NoStop}%
\bibitem [{\citenamefont {Zhou}(2021)}]{Zhou2021}%
  \BibitemOpen
  \bibfield  {author} {\bibinfo {author} {\bibfnamefont {Y.}~\bibnamefont
  {Zhou}},\ }\href
  {https://doi.org/https://doi.org/10.1016/j.physrep.2021.07.001} {\bibfield
  {journal} {\bibinfo  {journal} {Phys. Rep.}\ }\textbf {\bibinfo {volume}
  {935}}, \ \bibinfo {pages} {1} (\bibinfo {year} {2021})}
%,\ 
%\bibinfo {note}  {turbulence theories and statistical closure approaches}
\BibitemShut
  {NoStop}%
\bibitem [{\citenamefont {Richardson}(1922)}]{inRichardson1922}%
  \BibitemOpen
  \bibfield  {author} {\bibinfo {author} {\bibfnamefont {L.~F.}\ \bibnamefont
  {Richardson}},\ }\bibinfo {title} {Weather prediction by numerical
  processes}\ (\bibinfo  {publisher} {Cambridge University Press},\ \bibinfo
  {year} {1922})\ p.~\bibinfo {pages} {66}\BibitemShut {NoStop}%
\bibitem [{\citenamefont {Kolmogorov}(1941)}]{Kolmogorov1941a}%
  \BibitemOpen
  \bibfield  {author} {\bibinfo {author} {\bibfnamefont {A.~N.}\ \bibnamefont
  {Kolmogorov}},\ }\href@noop {} {\bibfield  {journal} {\bibinfo  {journal}
  {Dokl. Akad. Nauk SSSR}\ }\textbf {\bibinfo {volume} {30}},\ \bibinfo {pages}
  {301} (\bibinfo {year} {1941})},\ \bibinfo {note} {(reprinted in {\it Proc.
  R. Soc. Lond.} A {\bf 434}, 9)}\BibitemShut {NoStop}%
\bibitem [{\citenamefont {Frisch}(1995)}]{Frisch1995}%
  \BibitemOpen
  \bibfield  {author} {\bibinfo {author} {\bibfnamefont {U.}~\bibnamefont
  {Frisch}},\ }\href@noop {} {\emph {\bibinfo {title} {Turbulence: The Legacy
  of A. N. Kolmogorov}}}\ (\bibinfo  {publisher} {Cambridge University Press},\
  \bibinfo {year} {1995})\BibitemShut {NoStop}%
\bibitem [{\citenamefont {Onsager}(1949)}]{Onsager1949}%
  \BibitemOpen
  \bibfield  {author} {\bibinfo {author} {\bibfnamefont {L.}~\bibnamefont
  {Onsager}},\ }\href@noop {} {\bibfield  {journal} {\bibinfo  {journal} {Nuovo
  Cimento Suppl.}\ }\textbf {\bibinfo {volume} {6}},\ \bibinfo {pages}
  {279} (\bibinfo {year} {1949})}\BibitemShut {NoStop}%
\bibitem [{\citenamefont {Eyink}(2018)}]{Eyink2018}%
  \BibitemOpen
  \bibfield  {author} {\bibinfo {author} {\bibfnamefont {G.~L.}\ \bibnamefont
  {Eyink}},\ }\href@noop {} 
  %{\bibinfo {title} {Review of the Onsager ``idealturbuence'' theory}}
  \Eprint
  {https://arxiv.org/abs/1803.02223} {arXiv:1803.02223 [physics.flu-dyn]}
  (\bibinfo {year} {2018})
  \BibitemShut {NoStop}%
\bibitem [{\citenamefont {Ishihara}\ \emph {et~al.}(2016)\citenamefont
  {Ishihara}, \citenamefont {Morishita}, \citenamefont {Yokokawa},
  \citenamefont {Uno},\ and\ \citenamefont {Kaneda}}]{Ishiharaetal2016}%
  \BibitemOpen
  \bibfield  {author} {\bibinfo {author} {\bibfnamefont {T.}~\bibnamefont
  {Ishihara}}, \bibinfo {author} {\bibfnamefont {K.}~\bibnamefont {Morishita}},
  \bibinfo {author} {\bibfnamefont {M.}~\bibnamefont {Yokokawa}}, \bibinfo
  {author} {\bibfnamefont {A.}~\bibnamefont {Uno}},\ and\ \bibinfo {author}
  {\bibfnamefont {Y.}~\bibnamefont {Kaneda}},\ }\href@noop {} {\bibfield
  {journal} {\bibinfo  {journal} {Phys. Rev. Fluids}\ }\textbf {\bibinfo
  {volume} {1}},\ \bibinfo {pages} {082403(R)} (\bibinfo {year}
  {2016})}\BibitemShut {NoStop}%
\bibitem [{\citenamefont {\v{S}ver\'{a}k}(2017)}]{Sverak2017}%
  \BibitemOpen
  \bibfield  {author} {\bibinfo {author} {\bibfnamefont {V.}~\bibnamefont
  {\v{S}ver\'{a}k}},\ }in\ \href@noop {} {\emph {\bibinfo {booktitle}
  {Vector-Valued Partial Differential Equations and Applications}}},\ \bibinfo
  {editor} {edited by\ \bibinfo {editor} {\bibfnamefont {J.}~\bibnamefont
  {Ball}}\ and\ \bibinfo {editor} {\bibfnamefont {P.}~\bibnamefont
  {Marcellini}}}\ (\bibinfo  {publisher} {Springer},\ \bibinfo {year} {2017})\
  pp.\ \bibinfo {pages} {195--248}\BibitemShut {NoStop}%
\bibitem [{\citenamefont {Kaneda}(1987)}]{Kaneda1987}%
  \BibitemOpen
  \bibfield  {author} {\bibinfo {author} {\bibfnamefont {Y.}~\bibnamefont
  {Kaneda}},\ }\href@noop {} {\bibfield  {journal} {\bibinfo  {journal} {Phys.
  Fluids}\ }\textbf {\bibinfo {volume} {30}},\ \bibinfo {pages} {2672} (\bibinfo
  {year} {1987})}
%,\ \bibinfo {note} {erratum: see the reference in T.~Ishihara and Y.~Kaneda, Phys. Fluids, {\bf{13}} (2001) 544}
  \BibitemShut {NoStop}%
\bibitem [{\citenamefont {Kaneda}(2007)}]{Kaneda2007}%
  \BibitemOpen
  \bibfield  {author} {\bibinfo {author} {\bibfnamefont {Y.}~\bibnamefont
  {Kaneda}},\ }\href@noop {} {\bibfield  {journal} {\bibinfo  {journal} {Fluid
  Dyn. Res.}\ }\textbf {\bibinfo {volume} {39}},\ \bibinfo {pages}
  {526} (\bibinfo {year} {2007})}\BibitemShut {NoStop}%
\bibitem [{\citenamefont {Ishihara}\ and\ \citenamefont
  {Kaneda}(2001)}]{IshiharaKaneda2001}%
  \BibitemOpen
  \bibfield  {author} {\bibinfo {author} {\bibfnamefont {T.}~\bibnamefont
  {Ishihara}}\ and\ \bibinfo {author} {\bibfnamefont {Y.}~\bibnamefont
  {Kaneda}},\ }\href@noop {} {\bibfield  {journal} {\bibinfo  {journal}
  {Phys. Fluids}\ }\textbf {\bibinfo {volume} {13}},\ \bibinfo {pages}
  {544} (\bibinfo {year} {2001})}\BibitemShut {NoStop}%
\bibitem [{\citenamefont {Leslie}(1973)}]{inLeslie1973_7}%
  \BibitemOpen
  \bibfield  {author} {\bibinfo {author} {\bibfnamefont {D.~C.}\ \bibnamefont
  {Leslie}},\ }\bibinfo {title} {Developments in the theory of turbulence}\
  (\bibinfo  {publisher} {Clarendon Press, Oxford},\ \bibinfo {year} {1973})\
  Chap.~\bibinfo {chapter} {7}\BibitemShut {NoStop}%
\bibitem [{\citenamefont {Qian}(1996)}]{Qian1996}%
  \BibitemOpen
  \bibfield  {author} {\bibinfo {author} {\bibfnamefont {J.}~\bibnamefont
  {Qian}},\ }\href@noop {} {\bibfield  {journal} {\bibinfo  {journal} {J. 
  Phys. A: Math. Gen.}\ }\textbf {\bibinfo {volume} {29}},\
  \bibinfo {pages} {1305} (\bibinfo {year} {1996})}\BibitemShut {NoStop}%
\bibitem [{\citenamefont {Boffetta}\ and\ \citenamefont
  {Ecke}(2012)}]{BoffettaEcke2012}%
  \BibitemOpen
  \bibfield  {author} {\bibinfo {author} {\bibfnamefont {G.}~\bibnamefont
  {Boffetta}}\ and\ \bibinfo {author} {\bibfnamefont {R.~E.}\ \bibnamefont
  {Ecke}},\ }\href@noop {} {\bibfield  {journal} {\bibinfo  {journal} {Ann. 
  Rev. Fluid Mech.}\ }\textbf {\bibinfo {volume} {44}},\ \bibinfo
  {pages} {427} (\bibinfo {year} {2012})}\BibitemShut {NoStop}%
\bibitem [{\citenamefont {Kaneda}\ and\ \citenamefont
  {Ishihara}(2001)}]{KanedaIshihara2001}%
  \BibitemOpen
  \bibfield  {author} {\bibinfo {author} {\bibfnamefont {Y.}~\bibnamefont
  {Kaneda}}\ and\ \bibinfo {author} {\bibfnamefont {T.}~\bibnamefont
  {Ishihara}},\ }\href@noop {} {\bibfield  {journal} {\bibinfo  {journal}
  {Phys. Fluids}\ }\textbf {\bibinfo {volume} {13}},\ \bibinfo {pages} {1431}
  (\bibinfo {year} {2001})}\BibitemShut {NoStop}%
\end{thebibliography}
\end{document}